\documentclass[preprint,review, 12pt]{elsarticle}
 
\usepackage{amssymb}

\usepackage{lineno}
\usepackage{graphicx}
\usepackage{subfigure}
\usepackage{multirow}
\usepackage{amsmath}
\usepackage{amssymb}
\usepackage{ulem}
\usepackage{verbatim}
\usepackage{upgreek}
\journal{NIMA}
\usepackage{xcolor}

\usepackage{xspace} 
\makeatletter
\ifcase \@ptsize \relax
  \newcommand{\miniscule}{\@setfontsize\miniscule{4}{5}}
\or
  \newcommand{\miniscule}{\@setfontsize\miniscule{5}{6}}
\or
  \newcommand{\miniscule}{\@setfontsize\miniscule{5}{6}}
\fi
\makeatother
\DeclareRobustCommand{\optbar}[1]{\shortstack{{\miniscule (\rule[.5ex]{0.75em}{.18mm})}
  \\ [-.7ex] $#1$}}
\def\porpbar    {\kern 0.18em \optbar{\kern -0.18em p}{}\xspace}
 \def\Ppi         {\ensuremath{\pi}\xspace}         
 \def\PK      {\ensuremath{K}\xspace}            
\def\pion   {{\ensuremath{\Ppi}}\xspace}
\def\kaon    {{\ensuremath{\PK}}\xspace}
\def\pipm   {{\ensuremath{\pion^\pm}}\xspace}
\def\Kpm     {{\ensuremath{\kaon^\pm}}\xspace}

\begin{document}

\begin{frontmatter}

\title{\boldmath Requirement analysis for dE/dx measurement and PID performance at the CEPC baseline detector}
\author[label1,label2]{Y.Zhu}
\author[label1,label2]{S.Chen}
\author[label1,label2]{H.Cui}
\author[label1,label2]{M.Ruan\corref{cor1}}
\ead{ruanmq@ihep.ac.cn}

\address[label1]{Institute of High Energy Physics, Chinese Academy of Sciences, 19B Yuquan Road, Shijingshan District, Beijing 100049, China}
\address[label2]{University of Chinese Academy of Sciences, 19A Yuquan Road, Shijingshan District, Beijing 100049, China}

\begin{abstract}

The Circular Electron-Positron Collider (CEPC) can be operated not only as a Higgs factory but also as a Z-boson factory, offering great opportunities for flavor physics studies where Particle Identification (PID) is critical.
The baseline detector of the CEPC could record TOF and dE/dx information that can be used to distinguish particles of different species.
We quantify the physics requirements and detector performance using physics benchmark analyzes with full simulation.
We conclude that at the benchmark TOF performance of $50\,$ps, the dE/dx resolution should be better than 3\% for incident particles in the barrel region with a relevant momentum larger than $2\, $GeV/c.
This performance leads to an efficiency/purity for $K^{\pm}$ identification of 97\%/96\%, for $D^0\to \pi^+K^-$ reconstruction of 68.19\%/89.05\%, and for $\phi\to K^+K^-$ reconstruction of 82.26\%/77.70\%, providing solid support for relevant CEPC flavor physics measurements.

\end{abstract}

\begin{keyword}
CEPC \sep TPC \sep PID 
\end{keyword}

\end{frontmatter}

\section{Introduction}

The Circular Electron Positron Collider (CEPC) \cite{CEPCStudyGroup:2018ghi} is a large-scale collider facility proposed after the discovery of Higgs boson in 2012.
It is designed with a circumference of $100\,$km with two interaction points.
It can operate at multiple center-of-mass energies, including $240\,$GeV as a Higgs factory, $160\,$GeV for the $W^+W^-$ threshold scan, $91\,GeV$ as a Z factory, and $360\,$GeV for the $t\bar{t}$ threshold scan.
Table~\ref{CECPope} in Ref.~\cite{CEPCPhysicsStudyGroup:2022uwl} summarizes its baseline operating scheme and the corresponding boson yields.
In the future, it can be upgraded to a proton-proton collider to directly scan new physics signals at a center-of-mass energy of about $75\,$TeV.

\begin{table}[htbp]
\centering
\caption{\label{CECPope}The operation scheme of the CEPC, including the center-of-mass energy,  the instantaneous luminosity, the total integrated luminosity, and the event yields.}
\smallskip
\begin{tabular}{ccccc}
\hline
Operation mode    & Z factory     & WW     & Higgs factory    & $t\bar{t}$ \\
\hline
$\sqrt{s}$ (GeV) & 91.2   & 160     & 240      & 360 \\
Run time (year)  & 2 & 1 & 10 & 5 \\
Instantaneous luminosity      &   \multirow{2}{*}{191.7}    &   \multirow{2}{*}{26.6}    &   \multirow{2}{*}{8.3}   &  \multirow{2}{*}{0.83} \\
($10^{34}cm^{-2} s^{-1}$, per IP)  & & & & \\
Integrated luminosity      &  \multirow{2}{*}{100}    &   \multirow{2}{*}{6}    &   \multirow{2}{*}{20}   &  \multirow{2}{*}{1} \\
($ab^{-1}$, 2 IPs) & & & & \\
Event yields   &  $3\times 10^{12}$  & $1\times 10^8$ & $4\times 10^6$ & $5 \times 10^{5}$ \\
\hline
\end{tabular}
\end{table}

The main scientific objective of the CEPC is the precise measurement of the Higgs properties. 
When CEPC operates as a Z-boson factory, trillions of generated $Z\to q\bar{q}$ events can provide a great opportunity for measuring flavor physics, where particle identification (PID) is essential.

The baseline CEPC detector uses a TPC as the main tracker, which can record the dE/dx or even the cluster count dN/dx \cite{Cuna:2021sho, Chiarello:2019eny} information from tracks.
In addition, the CEPC detector is proposed to have TOF capability with cluster-level precision of $50\,$ps.
The TOF and dE/dx information depend on the particle species and its momentum, and enable \Kpm/\pipm/\porpbar identification.
In this paper, the dE/dx performance of the CEPC TPC is investigated based on Monte Carlo (MC) simulation, which induces the intrinsic dE/dx resolution.
In real experiments, both detector effects and imperfect calibration can deteriorate the dE/dx resolution.
Based on the potential degradation in previous TPCs by comparing their experimental achievements with the corresponding intrinsic dE/dx resolutions obtained from MC simulation,
we quantify the performance requirements of dE/dx using full simulated Z-pole samples corresponding to the Z-pole operation of the CEPC.

This article is organized as follows. 
Section~\ref{Det} introduces the detector, software, and samples used in this analysis.
Section~\ref{sec:PID} describes the measurement of TOF and dE/dx and investigates the \Kpm/\pipm/\porpbar separation performance.
Section~\ref{sec:K} presents the performance of $K^{\pm}$ identification, $D^0\to \pi^+K^-$ and $\phi\to K^+K^-$ reconstruction efficiency/purity using the inclusive hadronic Z-pole samples.
A brief conclusion is given in Section~\ref{sec:Con}.

\section{The reference detector, software, and samples}
\label{Det}

\begin{figure}[htbp]
\centering
\includegraphics[width=.55\textwidth]{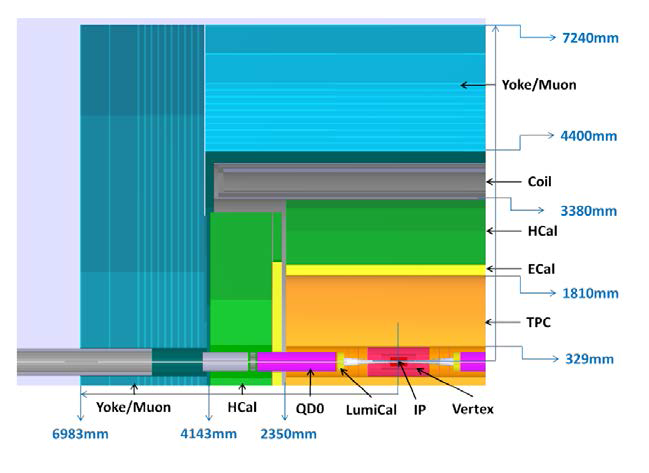}
\caption{\label{det}The structure of the baseline CEPC detector design.}
\end{figure}

\begin{figure}[htbp]
\centering
\includegraphics[width=.5\textwidth]{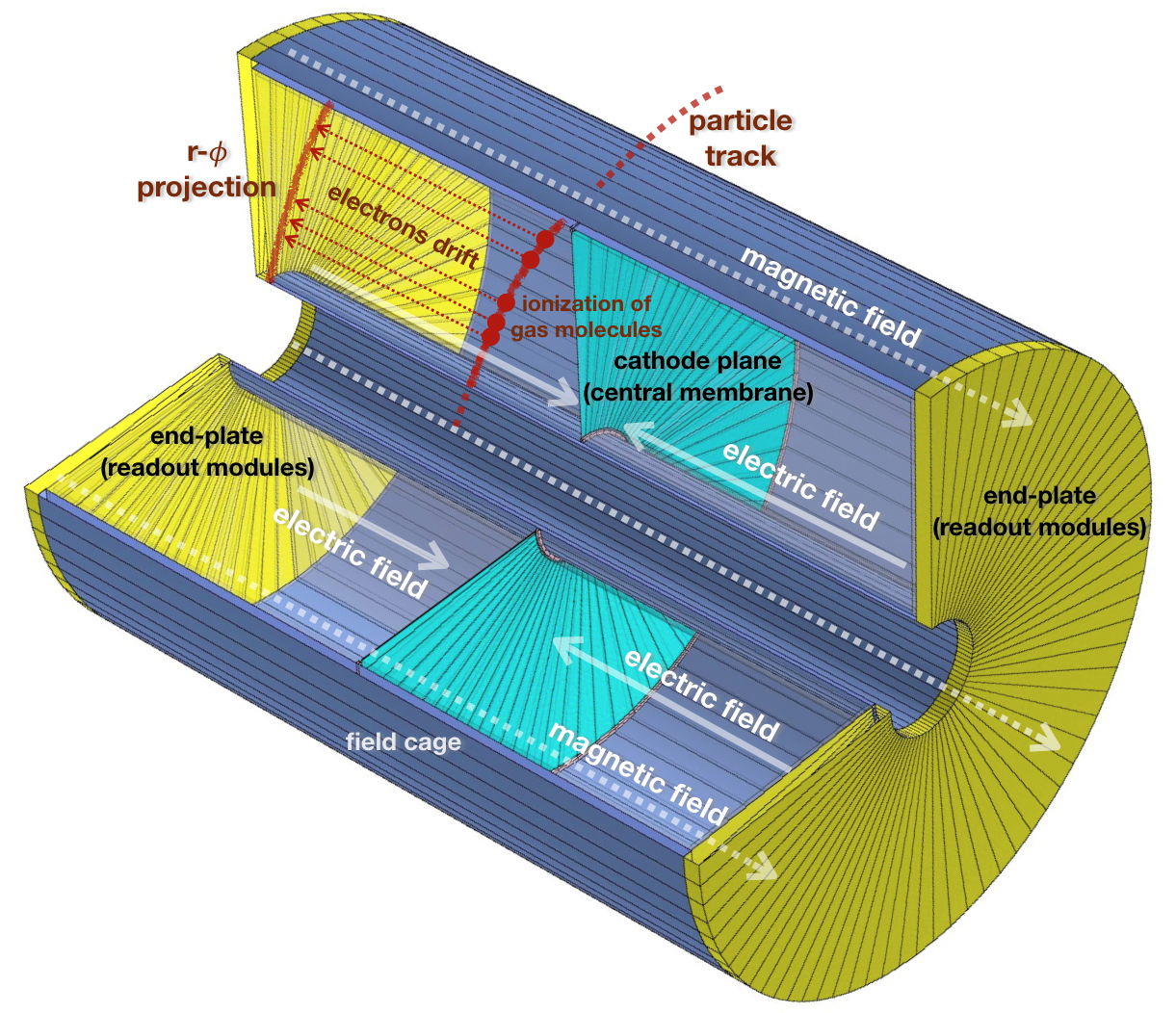}
\caption{\label{TPC}The structure of the TPC in the CEPC baseline detector design. This TPC is a cylindrical gas detector with an axial electric field formed between the end-plates (yellow) and a central cathode plane/membrane (light blue). The cylindrical walls of the volume form the electric field cage (dark blue). Gas ionization electrons due to charged particles drift to the end-plates where they are collected by readout modules (yellow).}
\end{figure}

The layout of the baseline CEPC detector design is shown in Fig.~\ref{det}.
It is designed following the particle flow principle \cite{arbor}, which emphasizes on the separation of final state particles and measures each final state particle in the most suited subdetector.
From innermost part of the detector layout to outermost part, the baseline concept consists of a silicon pixel vertex detector, a silicon inner tracker, a TPC surrounded by a silicon external tracker, a silicon-tungsten sampling ECAL, a steel-Glass Resistive Plate Chambers sampling Hadronic Calorimeter (HCAL), a 3-Tesla superconducting solenoid, and a flux return yoke embedded with a muon detector.
The PID performance in this paper is derived by dE/dx and TOF information collected by TPC and ECAL of baseline CEPC detector.
The TOF information recorded by ECAL can reach a time resolution of $50\, $ps \cite{TOF}, which can be achieved with modern silicon sensing technologies.
Figure~\ref{TPC} shows the structure of the TPC, which is located between the inner and outer silicon trackers and contains 220 radial layers with $6\, $mm increments.
This TPC has a cylindrical drift volume that provides a very homogeneous electric field.
The drift volume is filled with gases combining Ar/CF$_4$/C$_4$H$_{10}$ at atmospheric pressure with ratios of 95\%/3\%/2\%.
A detailed description of the TPC can be found in Ref.~\cite{CEPCStudyGroup:2018ghi}.

A baseline reconstruction software chain has been developed to evaluate the physics performance of the CEPC baseline detector.
The data flow of CEPC baseline software starts from the event generators of Whizard~\cite{Stienemeier:2021cse, Moretti:2001zz} and Pythia~\cite{Bierlich:2022pfr}.
The detector geometry is implemented into the MokkaPlus, a GEANT4~\cite{geant4} based full simulation module.
The MokkaPlus calculates the energy deposition in the detector sensitive volumes and creates simulated hits.
For each sub-detector, the digitization module converts the simulated hits into digitized hits by convoluting the corresponding sub-detector responses.
The reconstruction modules include  tracking,  Particle Flow, and  high-level reconstruction algorithms.
The digitized tracker hits are reconstructed into tracks via the tracking algorithm.
The Particle Flow algorithm, Arbor, reads the reconstructed tracks and the calorimeter hits to build  particle events.
High-level reconstruction algorithms reconstruct composite physics objects and identify the flavor of the jets.

Using the CEPC baseline detector geometry and software chain, we simulated $6.313\times 10^{6}$  inclusive hadronic Z-pole events.
These events include all the different quark flavors according to the SM predictions, whose details are shown in Table~\ref{zpole}.

\begin{table}[htbp]
\centering
\caption{\label{zpole}The branching ratio and the number of simulated events of the Z-pole samples.}
\smallskip
\begin{tabular}{cc|c}
\hline
Process & $\mathcal{B}$  & Sample used \\
\hline
$Z\to u\overline{u}$ & 11.17\%   & \multirow{5}{*}{ $6.313\times 10^{6}$}\\
\cline{1-2}
$Z\to d\overline{d}$ & 15.84\% &  \\
\cline{1-2}
$Z\to s\overline{s}$ & 15.84\%  &  \\
\cline{1-2}
$Z\to c\overline{c}$ & 12.03\%  &  \\
\cline{1-2}
$Z\to b\overline{b}$ & 15.12\%  &  \\
\hline
\end{tabular}
\end{table}

\section{PID performance using TOF and dE/dx}
\label{sec:PID}

This section presents the measurements of TOF and dE/dx information by ECAL and TPC, respectively, and the corresponding \Kpm/\pipm/\porpbar separation performance.
The separation power is defined as 
\begin{equation}
S_{AB} = \frac{|O_A - O_B|}{\sqrt{\sigma_{A}^2 + \sigma_{B}^2}},
\end{equation}
where $O_{A(B)}$ is the TOF or dE/dx of particle A(B) and $\sigma_{A(B)}$ is the TOF resolution or dE/dx resolution of particle A(B).
There are also other ways to define the separation power using different conventions.
To be consistent with the definition in the previous paper~\cite{An:2018jtk}, we use the definition above.

\subsection{PID performance using TOF}
The TOF of a given particle can be calculated based on its momentum and its flight distance along its trajectory.
In magnetic field, the trajectory of a charged particle would be a helix that can interact with the barrel region or the end-cap region of ECAL, depending on the four-momentum of the incident particle.
The inner radius of ECAL is represented by R and the radius of the helix is derived as

\begin{equation}
\label{eq:tof}
\begin{split}
r = \frac{1000\cdot P_t}{0.3\cdot B\cdot q} (mm),
\end{split}
\end{equation}
where $P_t$, $q$, and B are the transverse momentum in GeV/c, the charge of the incident particle, and the magnetic field strength in the TPC in T, respectively.
Once the incident particle interacts with the barrel of the ECAL, the length of the trajectory of the incident particle can be described by $arc = 2\pi r \cdot \frac{\phi}{2\pi}$, and the corresponding arc angle is represented by $\phi = 2 \cdot arcsin(\frac{R}{2r})$.
The velocity of the particle perpendicular/parallel to the B-field is represented by $v_t$/$v_l$, and the length between the Interaction Point (IP) and the ECAL end cap is represented by $L$.
At the CEPC, $L = 2.35\, $m and $R = 1.8\, $m.
If $2\cdot r$ $>$ R \& $L/v_l > arc/v_t$, the helix would interact with the barrel of ECAL, $TOF = arc/v_t$.
Otherwise, the helix would interact with the end cap of ECAL, $TOF = L/v_l$.

\begin{figure}[htbp]
\centering
\includegraphics[width=.5\textwidth]{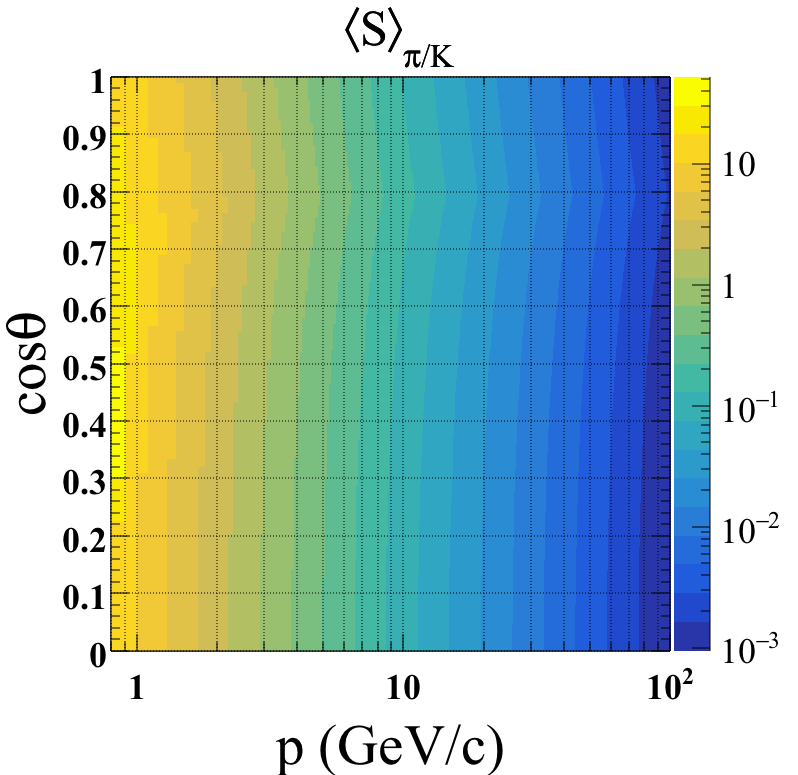}
\caption{\label{2DTOF} The $K^{\pm}/\pi^{\pm}$ separation power as a function of momentum and cosine polar angle including TOF information.}
\end{figure}

The $K^{\pm}/\pi^{\pm}$ separation power as a function of momentum and polar angle $\theta$ is shown in Fig.~\ref{2DTOF}.
The separation power is peaking around $cos\theta = 0.8$ corresponding to the maximum drift length.

\subsection{PID performance using dE/dx}

\begin{figure}[htbp]
\centering
\subfigure[]{ \label{fig:cdEdx_KPi}
		\includegraphics[width=0.45\textwidth]{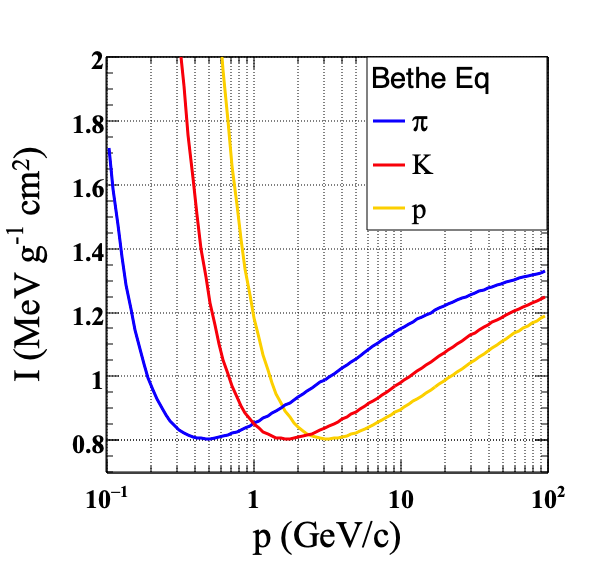}}
	\subfigure[]{ \label{fig:cDeltadEdx_kpi}
		\includegraphics[width=0.45\textwidth]{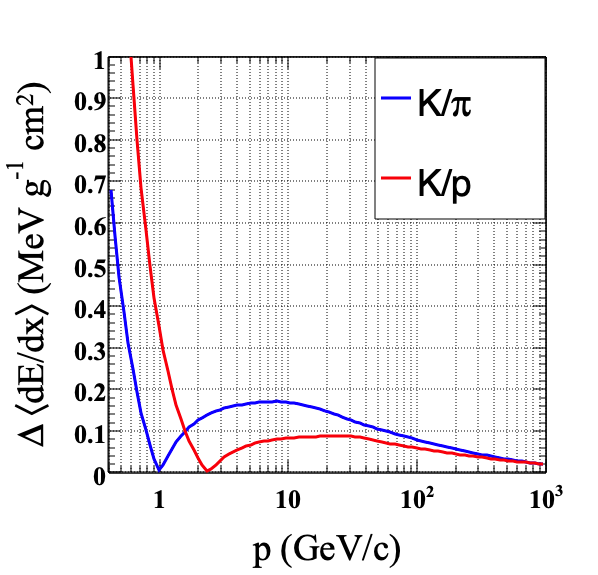}}
\caption{\label{dEdx}The distribution of $I$ as a function of momentum for \Kpm/\pipm/\porpbar (a) and the absolute difference of $I$ for \Kpm/\pipm and \Kpm/\porpbar (b).}
\end{figure}

The measured dE/dx of a track follows a Landau distribution with a large tail caused by high-energy $\delta$-electrons.
The average dE/dx for a track, denoted $I$, is estimated using the usual ``truncated mean'' method, where the truncation ratio of 90\% is determined to obtain the optimal dE/dx resolution, denoted $\sigma_I$.
For \Kpm/\pipm/\porpbar, the distributions of $I$ as a function of momentum are shown in Fig.~\ref{fig:cdEdx_KPi}.
The distributions of the absolute difference of $I$ for \Kpm/\pipm and \Kpm/\porpbar are shown in Fig.~\ref{fig:cDeltadEdx_kpi}.
The $K^{\pm}$ and $\pi^{\pm}$ with the same dE/dx at $1\, $GeV/c ($K^{\pm}$ and \porpbar at $2.5\, $GeV/c)  have no separation power.

\begin{figure}[htbp]
\centering
\subfigure[]{ \label{fig:PhotonAngle}
		\includegraphics[width=0.31\textwidth]{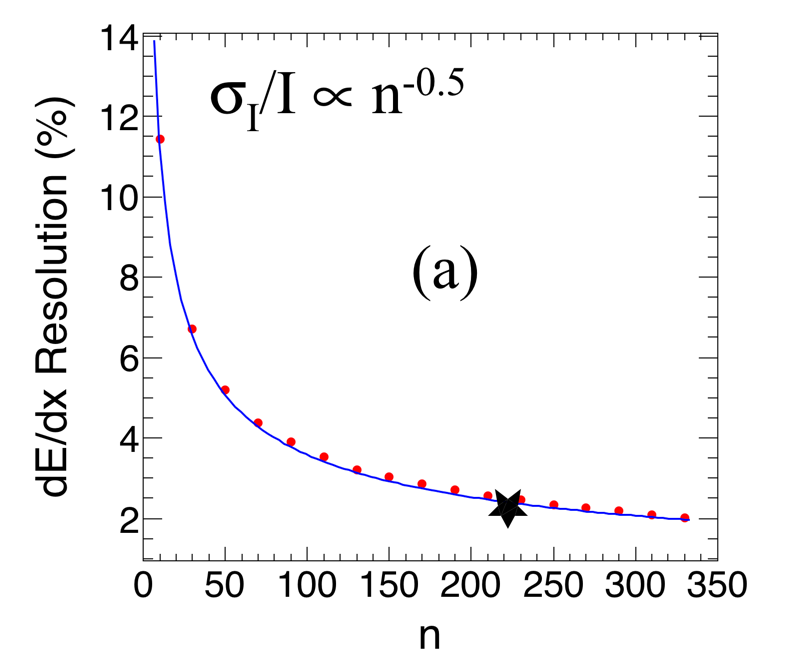}}
	\subfigure[]{ \label{fig:PhotonSeparation}
		\includegraphics[width=0.31\textwidth]{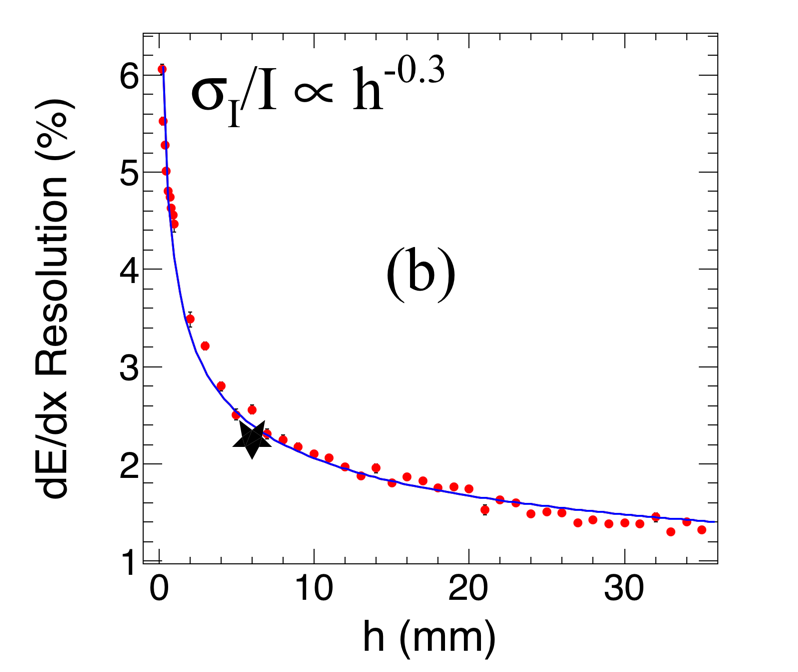}}
  	\subfigure[]{ \label{fig:dedx2rho}
		\includegraphics[width=0.31\textwidth]{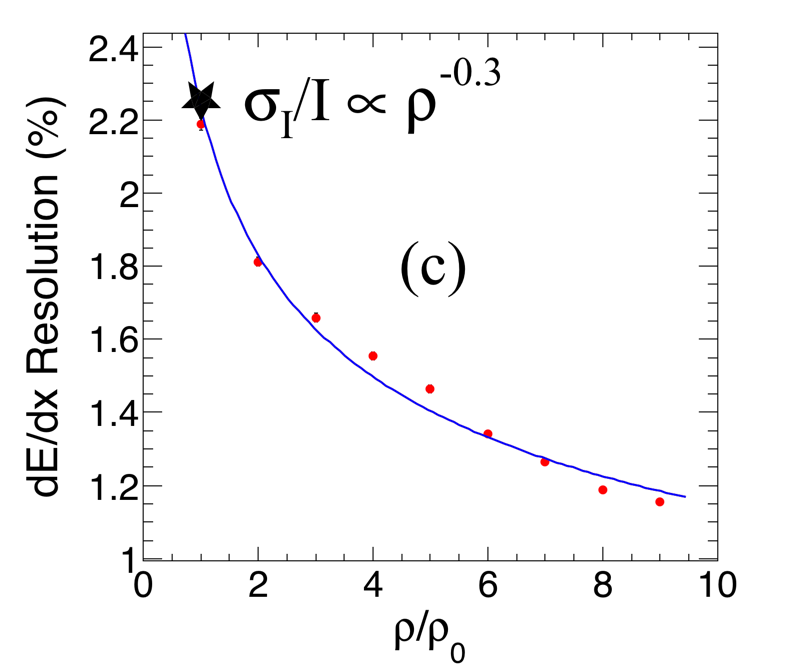}}
  \\
  	\subfigure[]{ \label{fig:dedx2bg}
		\includegraphics[width=0.31\textwidth]{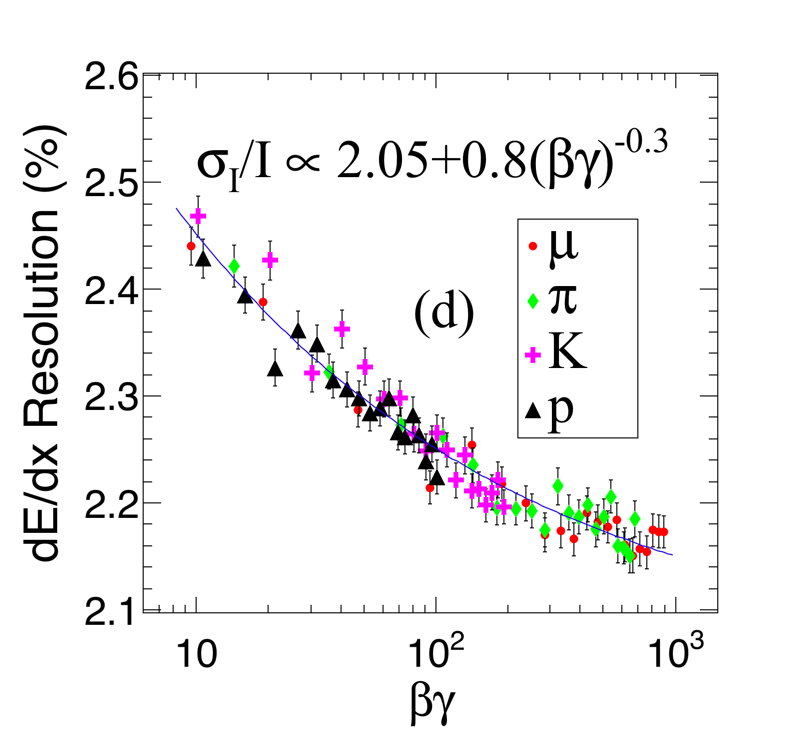}}
  	\subfigure[]{ \label{fig:dedx2Cos}
		\includegraphics[width=0.31\textwidth]{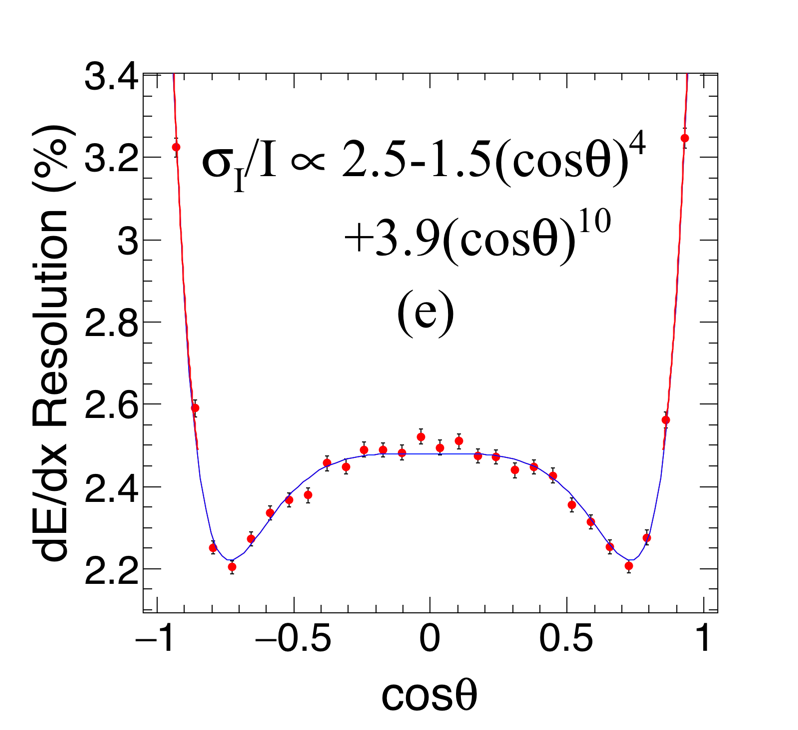}}
	\caption{ \label{differencial}The intrinsic dE/dx resolution versus the number of pad rings (a), the pad height along the radial direction (b), the ratio of gas density $\rho$ over the default gas density $\rho_0$ (equivalent to the ratio of corresponding pressures) (c), the relativistic velocity $\beta\gamma$ (d) and $cos\theta$ (e) of the ionizing particle. The default working point, where $n = 222$, $h = 6\,$mm, and $\rho = \rho_0 = 1.73\,$mg/cm$^3$, is indicated with a solid star symbol. Solid lines represent the fit projections. The dE/dx resolution in the end-cap region is described by two red lines in (e). }
\end{figure}

The predictive power of the dE/dx method depends on precise knowledge of the dE/dx energy loss parameterization and its resolution. 
The resolution induced by the design of the TPC and the momentum of the incident particle, including the number of pad rings $n$, the pad height along the radial direction $h$, the density of the working gas $\rho$, the relativistic velocity $\beta\gamma$, and the polar angle $\theta$ of the particle trajectory, is named ``intrinsic dE/dx resolution''.
While in real experiments, the dE/dx resolution could be deteriorated by the detector effects arising in the processes of electron drift, signal amplification and readout in TPC.
We name the dE/dx resolution in real experiments as ``actual dE/dx resolution".

\begin{figure}[htbp]
\centering
\subfigure[]{ \label{fig:cKsigma}
		\includegraphics[width=0.45\textwidth]{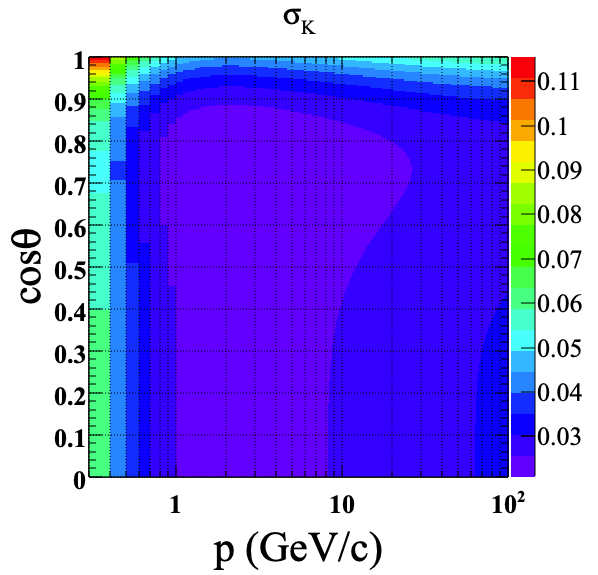}}
  \\
	\subfigure[]{ \label{fig:cPisigma}
		\includegraphics[width=0.45\textwidth]{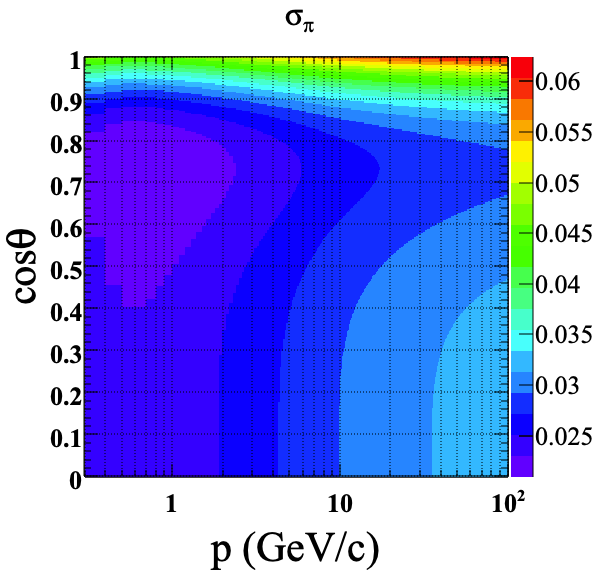}}
  \subfigure[]{ \label{fig:cPsigma}
		\includegraphics[width=0.45\textwidth]{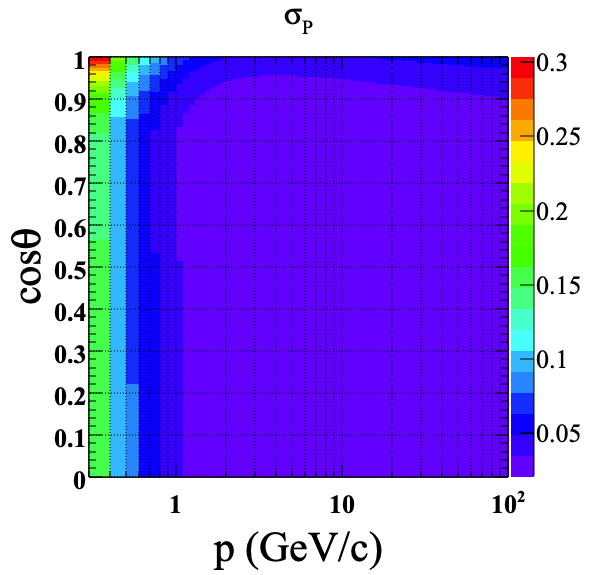}}
  \caption{\label{resokaon}The distributions of dE/dx resolution as a function of momentum and cosine polar angle for $K^{\pm}$ (a), $\pi^{\pm}$ (b), and \porpbar (c).}
\end{figure}

\begin{figure}[htbp]
\centering
	\subfigure[]{ \label{fig:piK}
		\includegraphics[width=0.45\textwidth]{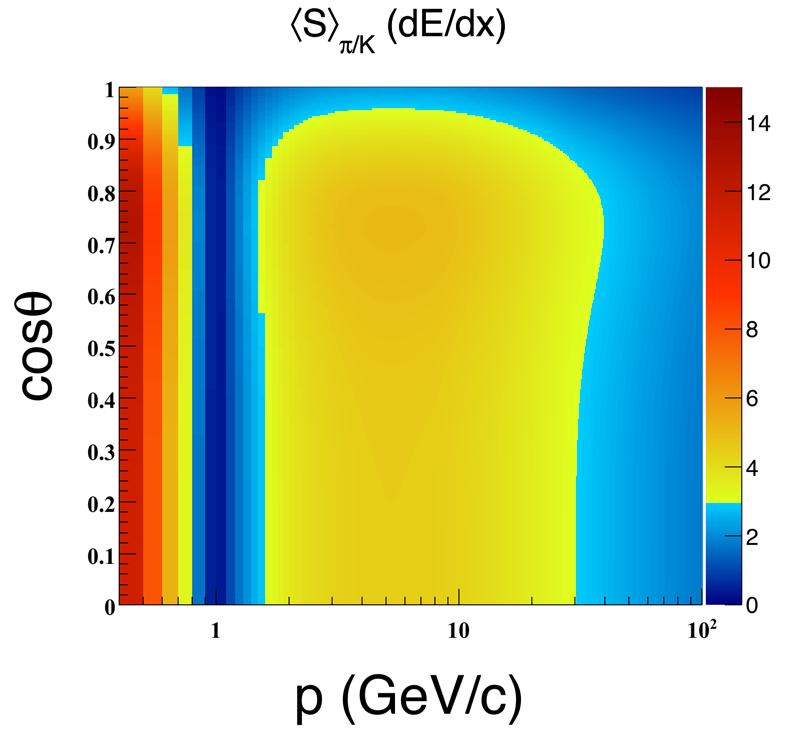}}
  \subfigure[]{ \label{fig:pK}
		\includegraphics[width=0.45\textwidth]{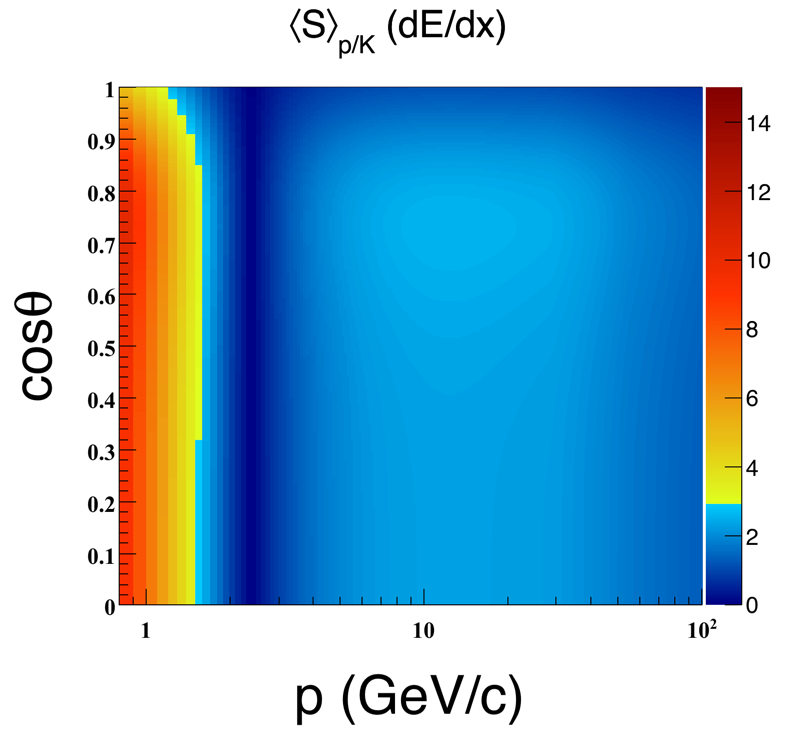}}
\caption{\label{cAn}Separation power as a function of momentum and cosine polar angle for \Kpm/\pipm (a) and \Kpm/\porpbar (b).}
\end{figure}

The differential of the dE/dx resolution, denoted $\sigma_I/I$, to the parameters of the TPC and the incident particle is shown in Fig.~\ref{differencial}.
The default values for the parameters of the TPC at the baseline CEPC detector are $n = 222$, $h = 6\,$mm, and $\rho = \rho_0 = 1.73\,$mg/cm$^3$, which are represented by the solid star markers in Fig.~\ref{differencial}.
The dE/dx resolution is a power function of $n$/$h$/$\rho$/$\beta\gamma$, while dE/dx resolution is a constant in the barrel region and becomes better with the increase of particle drift distance.
However, due to the relatively low physics performance of the end cap, the  dE/dx resolution becomes worse in the end-cap region ($|cos\theta| > 0.85$), see Fig.~\ref{fig:dedx2Cos}.
It can be seen that dE/dx resolution for \Kpm/\pipm/\porpbar can be better than 2.5\% when $\beta\gamma > 10$ in the barrel region. 
In real experiments, the dE/dx resolution in the barrel region is pursued to be better than 3\%. This means that the degradation of the dE/dx resolution caused by the electronics and the detector effects must be controlled to be less than 1/5 of the intrinsic dE/dx resolution in the barrel region.
The parameterization of dE/dx resolution can be factorized as 

\begin{equation}
\label{reso}
\begin{split}
\frac{\sigma_I}{I} = \frac{13.5}{n^{0.5}\cdot(h\rho)^{0.3}}\left[2.05 + 0.8(\beta\gamma)^{-0.3}\right]  \\
 \times \left[ 2.5 - 1.5(cos\Theta)^4 + 3.9(cos\Theta)^{10}\right],
 \end{split}
\end{equation}
where $h$ and $\rho$ are in mm and mg/cm$^3$, respectively.
The distribution of dE/dx resolution as a function of momentum and polar angle for \Kpm/\pipm/\porpbar is shown in Fig.~\ref{resokaon}.
The separation power as a function of momentum and polar angle is shown in Fig.~\ref{fig:piK} for \Kpm/\pipm and Fig.~\ref{fig:pK} for \Kpm/\porpbar. The regions with separation power higher than 3 are shown in warm colors.

\begin{figure}[htbp]
\centering
	\subfigure[]{ \label{fig:ccom}
		\includegraphics[width=0.45\textwidth]{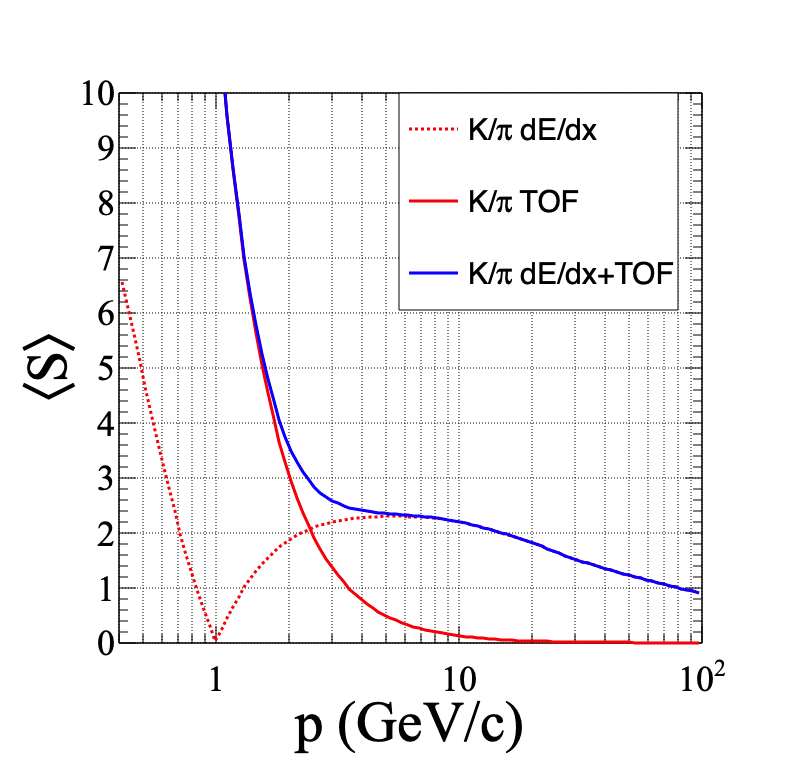}}
  \subfigure[]{ \label{fig:ccom_kp}
		\includegraphics[width=0.45\textwidth]{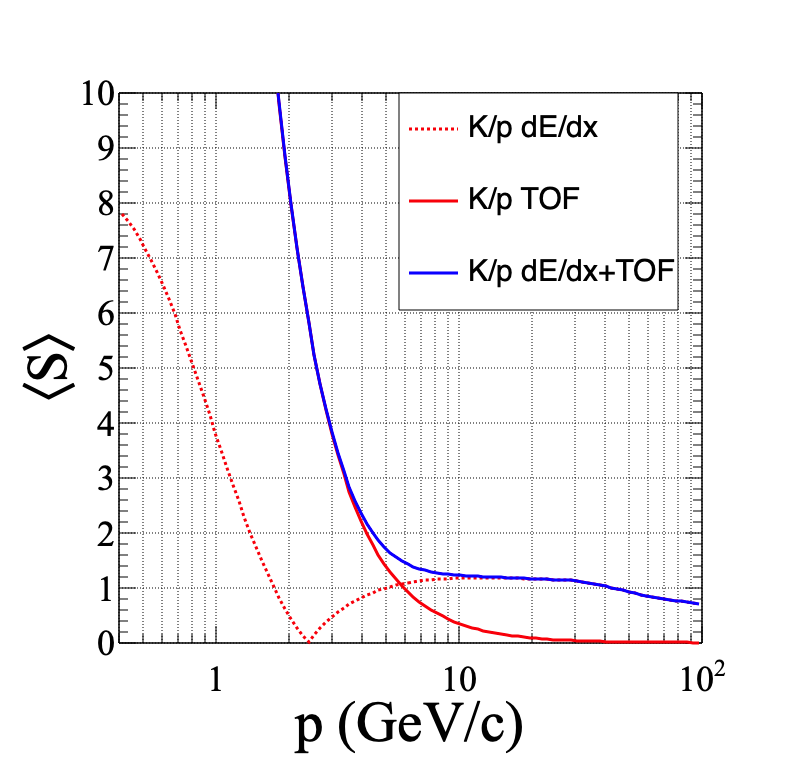}}
\caption{\label{ccom1D} The separation power as a function of momentum for \Kpm/\pipm (a) and \Kpm/\porpbar (b) at a polar angle of $\pi/4$.}
\end{figure}

\begin{figure*}[htbp]
\centering
	\subfigure[]{ \label{fig:piKcom}
		\includegraphics[width=0.45\textwidth]{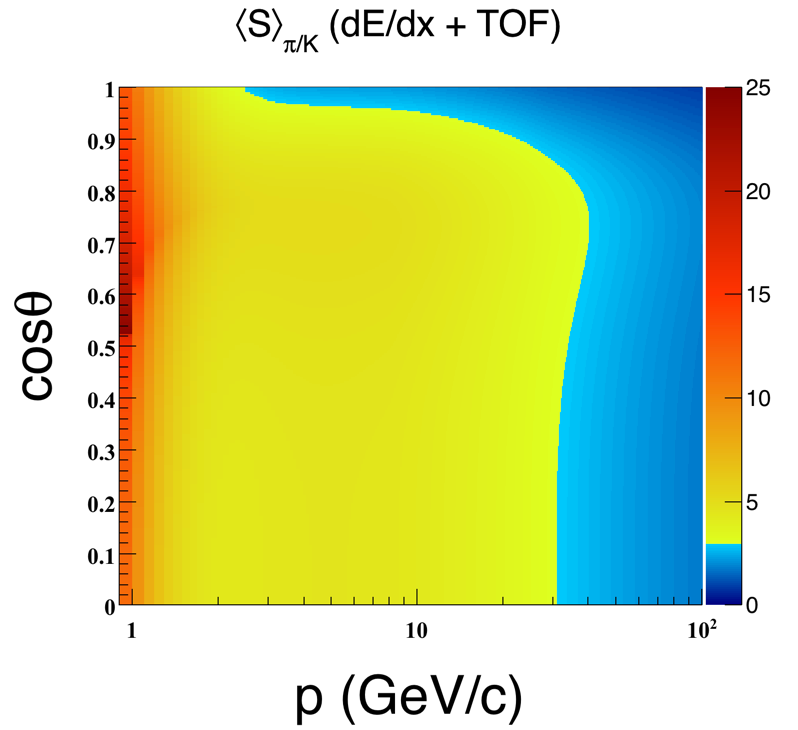}}
  \subfigure[]{ \label{fig:pKcom}
		\includegraphics[width=0.45\textwidth]{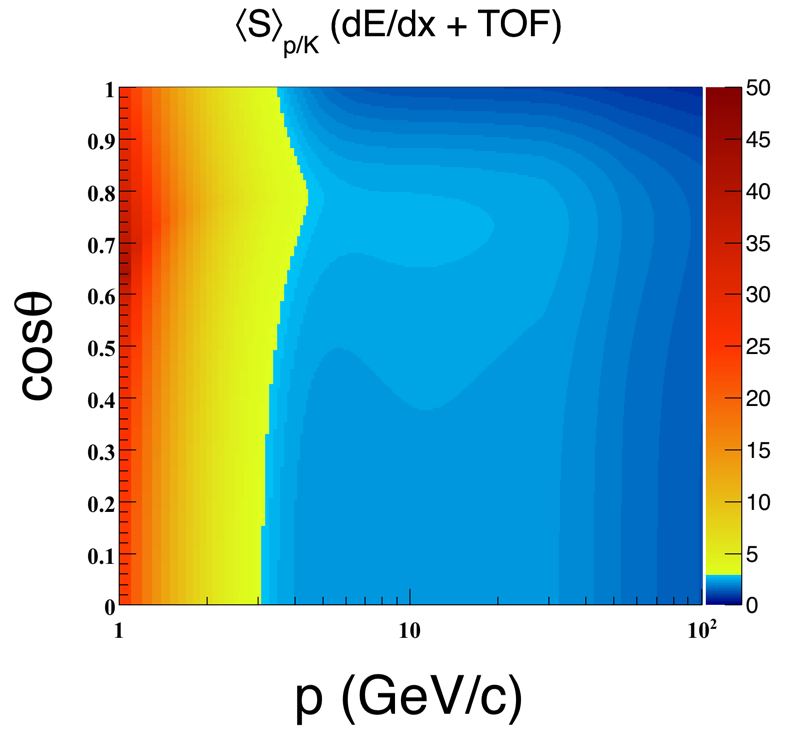}}
\caption{\label{ccom2D}Combine dE/dx and TOF, separation power as a function of momentum and cosine polar angle for \Kpm/\pipm (a) and \Kpm/\porpbar (b).
}
\end{figure*}

\subsection{PID performance combining both}

The above analyses investigate the separation power of \Kpm/\pipm/\porpbar with TOF and dE/dx information independently.
When the polar angle is $\pi/4$, the separation power as a function of momentum for \Kpm/\pipm and \Kpm/\porpbar is shown in Fig.~\ref{ccom1D}.
The TOF information can enhance the separation power of dE/dx information at about $1\, $GeV/c for \Kpm/\pipm separation and about $2.5\, $GeV/c for \Kpm/\porpbar separation.
After combining dE/dx and TOF, the separation power as a function of momentum and polar angle for \Kpm/\pipm and \Kpm/\porpbar is shown in Fig.~\ref{ccom2D}. The regions with separation power higher than 3 are shown in warm colors.

\section{PID evaluation}
\label{sec:K}

Since the separation power is convention dependent, we can also use the maximal efficiency times purity for \Kpm identification at a given sample, like the inclusive hadronic Z-pole samples listed in Table~\ref{zpole}, to study the PID performance of the baseline CEPC detector.
In addition, the identification efficiency and purity of any particle that decays into charged hadrons can also be used to quantify the PID performance.
Therefore, we extract the $D^0\to \pi^+K^-$ and $\phi\to K^+K^-$ reconstruction efficiency/purity using inclusive hadronic Z-pole samples.

\subsection{$K^{\pm}$ identification at the Z-pole}

\begin{figure}[htbp]
\centering
	\subfigure[]{ \label{fig:1088}
		\includegraphics[width=0.45\textwidth]{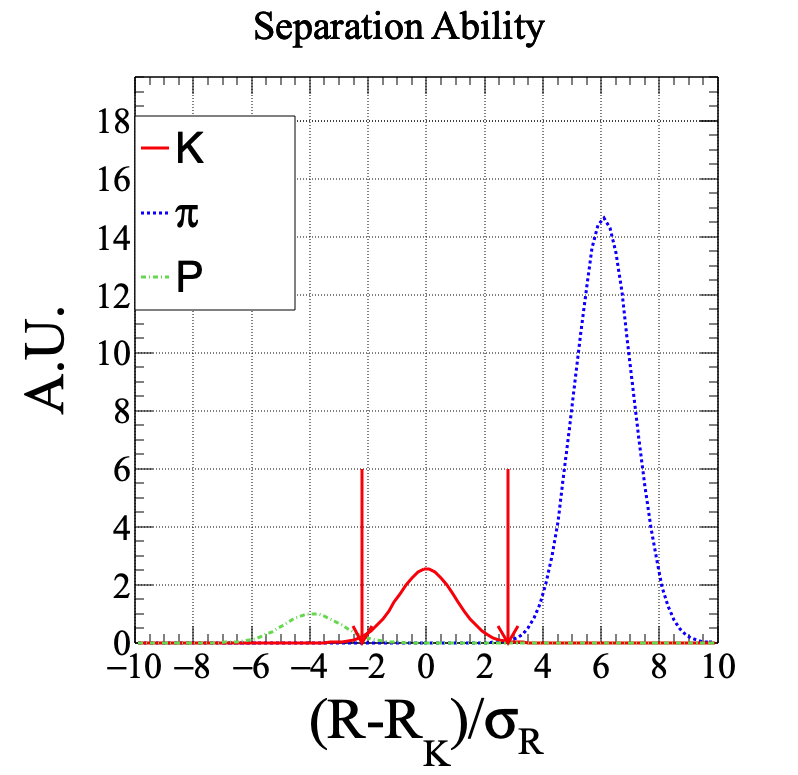}}
  \subfigure[]{ \label{fig:ceffpur}
		\includegraphics[width=0.45\textwidth]{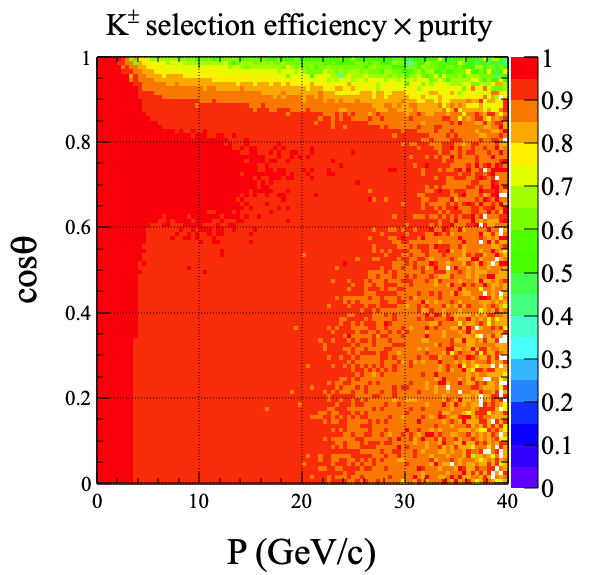}}
\caption{\label{sep28}By combining dE/dx and TOF information with intrinsic dE/dx resolution and $50\, $ps TOF resolution, the distribution of the variable $(R - R_K)/\sigma_R$ for a sample with \Kpm/\pipm/\porpbar at a momentum ranging from $12.0\, $GeV/c to $12.4\, $GeV/c and a $cos\theta$ ranging from 0.30 to 0.31 is shown in (a) and the performance of $K^{\pm}$ identification with maximum efficiency times purity at different momentum and cosine polar angle combinations is shown in (b).
}
\end{figure}

As demonstrated in Fig.~\ref{ccom2D}, the separation power has a strong dependence on the polar angle and momentum of the particles.
Therefore, for a given reference of momentum and polar angle range, we can optimize the cuts to select candidates with maximal efficiency times purity.
For example, Fig.~\ref{fig:1088} shows the distribution of the variable $(R - R_K)/\sigma_R$, where $R$ is the experimental measurement either by dE/dx alone or by the combination of dE/dx and TOF, $R_k$ is the expected value for the $K^{\pm}$ hypothesis, and $\sigma_R$ denotes the experimental resolution for \Kpm/\pipm/\porpbar in a given range of momentum and polar angle.
The two vertical lines indicate $K^{\pm}$ selection with maximal efficiency times purity.
The $K^{\pm}$ identification efficiency and purity are defined as

\begin{equation}
\begin{split}
\epsilon_K = \frac{N_{K\to K}}{N_K} 
\\
p_K = \frac{N_{K\to K}}{N_{K\to K} + N_{\pi \to K} + N_{p\to K}},
\end{split}
\end{equation}
where $N_K$ is the total number of $K^{\pm}$ produced, $N_{K\to K}$ is the number of $K^{\pm}$ correctly identified, and $N_{\pi(p)\to K}$ is the number of $\pi^{\pm}$(\porpbar) incorrectly identified as $K^{\pm}$.

By combining dE/dx and TOF with intrinsic dE/dx resolution and $50\, $ps TOF resolution, Fig.~\ref{fig:ceffpur} shows the maximal efficiency times purity of $K^{\pm}$ identification at different ranges of momentum and polar angle.
The overall efficiency and purity of $K^{\pm}$ identification is calculated as 
\begin{equation}
\begin{split}
overall\ efficiency = \frac{\sum_{bins} \epsilon_k \cdot N_k}{\sum_{bins} N_k} \\
overall\ purity = \frac{\sum_{bins} p_k \cdot N_k}{\sum_{bins} N_k},
\end{split}
\end{equation}
where $\sum_{bin}$ means summation over momentum in the range of 0 to $40\,$GeV/c and cosine polar angle in the range of 0 to 1, $\epsilon_k$, $p_k$ and $N_k$ represents the $K^{\pm}$ identification efficiency, purity and the number of $K^{\pm}$, respectively, in a given momentum and polar angle range.
With above convention, the overall efficiency and purity in Fig.~\ref{sep28} is 98.43\% and 97.89\%, respectively.

In reality, the dE/dx resolution is affected by detector effects and readout electronics.
With the definition $\sigma_{actual} = factor \cdot \sigma_{intrinsic}$, where $\sigma_{actual}$ is the actual dE/dx resolution and $\sigma_{intrinsic}$ is the intrinsic dE/dx resolution, the performance of $K^{\pm}$ identification under different factors with/without combining TOF information are shown in Table~\ref{kaonID}.
The factors are selected according to Table 1 in Ref.~\cite{An:2018jtk}, which shows several detectors from other high energy experiments have the actual dE/dx resolution decreased by 15\% to 50\% relative to the intrinsic ones.
The results in Table~\ref{kaonID} prove that worse dE/dx resolution leads to worse $K^{\pm}$ identification performance, while TOF information can improve $K^{\pm}$ identification performance on top of dE/dx information.
The Ref.~\cite{An:2018jtk} performs the same analysis, but the results differ from those in Table~\ref{kaonID}.
There are two reasons for this: First, the Ref.~\cite{An:2018jtk} selects candidates based on the intersections of the spectra shown in Fig.~\ref{fig:1088}, which is different from this analysis, and second, the momentum ranges for these two analyzes are different.

\begin{table}[htbp]
\centering
\caption{\label{kaonID}The $K^{\pm}$ identification performance with different factors, $\sigma_{actual} = factor \cdot \sigma_{intrinsic}$, with/without combination of TOF information at the Z-pole.}
\smallskip
\begin{tabular}{cccccc}
\hline
                                                  & factor             & 1.   & 1.2       &1.5      & 2.        \\
\hline
\multirow{2}{*}{dE/dx}						 & $\varepsilon_K$ (\%)                        &  95.97 & 94.09  &  91.19   &  87.09          \\
 				     & $purity_K$ (\%)                         &  81.56 & 78.17 & 71.85    &  61.28          \\
\hline
\multirow{2}{*}{dE/dx \& TOF}	 & $\varepsilon_K$ (\%)                 & 98.43  & 97.41 & 95.52    &  92.3         \\
 				         		 & $purity_K$ (\%)                          &  97.89 & 96.31   &  93.25   &   87.33        \\
\hline
\end{tabular}
\end{table}

\subsection{ $D^0\to \pi^+K^-$ reconstruction at the Z-pole}
\label{secD0}

$D^0$ mesons can be abundantly produced in the CEPC, especially in the Z-pole operating mode, and $D^0$ mesons are important research objects for the study of CP violation.
It is important to understand the reconstruction performance of $D^0$ in the baseline design of the CEPC detector.
Using the inclusive $Z\to q\overline{q}$ sample and under the condition of intrinsic dE/dx resolution, the reconstruction performance of $D^0\to \pi^+K^-$ in the baseline CEPC detector is investigated in this subsection. 

The method for reconstructing the $D^0\to \pi^+K^-$ is described below.
\begin{enumerate}
\item Match all pairs of reconstructed oppositely charged tracks.
\item To suppress the significant combinatorial backgrounds, the absolute mass difference between two oppositely charged tracks and $D^0$ is set to less than $10\, $MeV/c$^2$.
\item The impact parameter (IMP), which equivalent to the perpendicular distance between IP and the path of a track, is larger than $0.02\,$mm for each track.
\item The $\chi^2$ of the secondary vertex fit \cite{Suehara:2015ura} is less than 5.15.
\item The distance of the secondary vertex to IP is greater than $0.305\,$mm.
\item The identified PID of a positively charged particle is a $\pi^+$ and a negative $K^-$. 
\end{enumerate}
The efficiency and purity of the candidate $D^0$ selection are defined in Eq.~\ref{eq:effpur}.
To maximize the value of the efficiency times the purity of the signal selection, scans of the selection parameters were performed in the determination of the selection criteria.
The details of event selection can be found in Table~\ref{D0}.

\begin{equation}
\label{eq:effpur}
\begin{split}
\epsilon & = \frac{Number\ of\ correctly\ reconstructed\ candidate\ D^0}{Number\ of\ D^0\to \pi^+K^- decays}  \\
 p & = \frac{Number\ of\ correctly\ reconstructed\ candidate\ D^0}{Number\ of\ candidate\ D^0}
\end{split}
\end{equation}

\begin{table}[htbp]
\centering
\caption{\label{D0} The efficiency and purity of $D^0\to \pi^+K^-$ reconstructed using Z-pole sample. The second column ($\epsilon$) lists selection efficiency after each selection step, and the third column (p) lists the corresponding purity.}
\smallskip
\begin{tabular}{ccc}
\hline
						          & $\epsilon$ (\%) & p (\%)                     \\
\hline
$|mass - mass_{D0}| < 0.01\, $GeV/c$^2$	 & $90.39\pm 0.24$        &  $2.16\pm 0.07$            \\
IMP $> 0.02\, $mm					 & $79.12\pm 0.21$     &  $5.04\pm 0.11$                 \\
vertex fitted $\chi^2 <  5.15$		 & $72.62\pm 0.23$    &  $15.36\pm 0.18$             \\
dis of vertex to IP	$> 0.305\, $mm		&  $69.24\pm 0.24$     &  $28.41\pm 0.23$                 \\
PID                                 &  $68.19\pm 0.24$     &  $89.05\pm 0.16$                 \\
\hline
\end{tabular}
\end{table}

\begin{figure}[htbp]
\centering
\includegraphics[width=.5\textwidth]{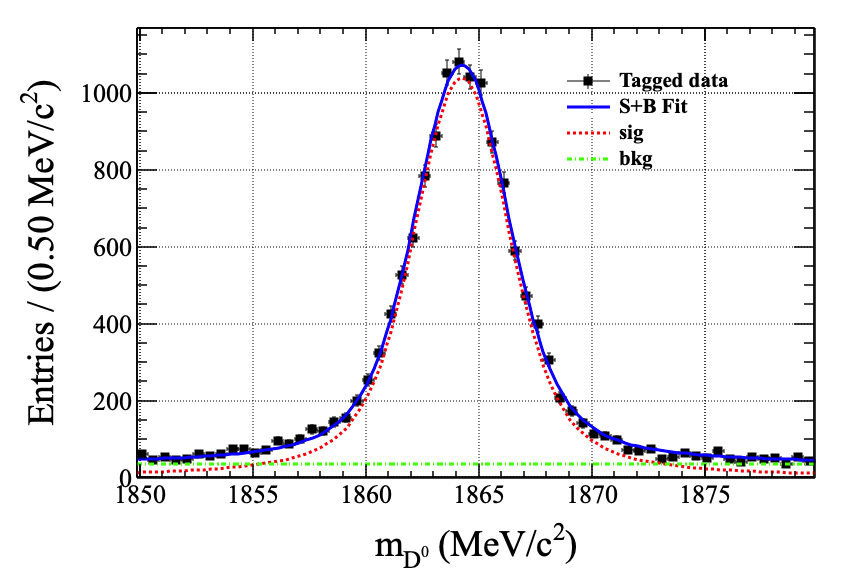}
\caption{\label{D0mass}Fit of the $D^0$ invariant mass of the track pair with inclusive hadronic Z-pole background included.}
\end{figure}

After applying the above selection criteria, the distribution of the invariant mass of the track pair is shown in Fig.~\ref{D0mass}.
The signal distribution is modeled with a Breit-Wigner theory model convolving a Gaussian detector response function, and the background distribution is modeled with a one order Chebyshev polynomial.
The mean value of fitted $D^0$ mass is $1864.259\pm 0.025\, $MeV/c$^2$.

\begin{figure}[htbp]
\centering
\includegraphics[width=.45\textwidth]{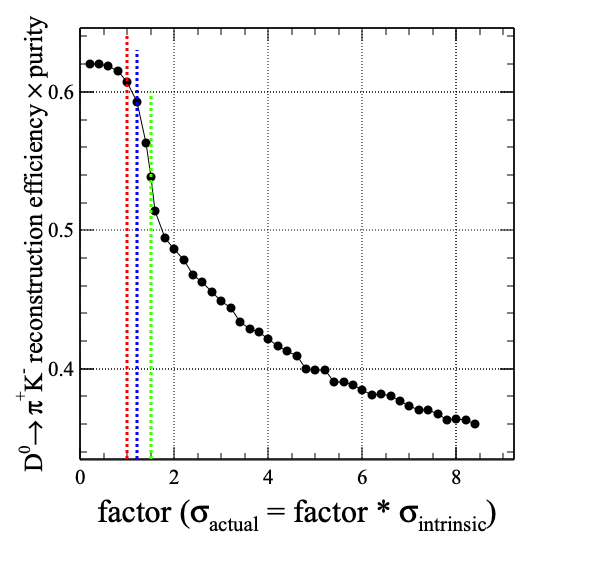}
\caption{\label{reso2}The distribution of $D^0\to \pi^+K^-$ reconstruction performance as a function of the factor defined in $\sigma_{actual} = factor \cdot \sigma_{intrinsic}$. The red/blue/green line corresponds to the 0\%/20\%/50\% degradation of dE/dx resolution.}
\end{figure}

$D^0\to \pi^+K^-$ reconstruction performance as a function of dE/dx resolution is shown in Fig.~\ref{reso2}, which proves that better dE/dx resolution leads to better $D^0\to \pi^+K^-$ reconstruction performance.
The red/blue/green line corresponds to 0\%/20\%/50\% degradation of dE/dx resolution.
It can be seen that 20\% degradation of dE/dx resolution would not significantly degrade the $D^0\to \pi^+K^-$ reconstruction performance.

\subsection{$\phi\to K^+K^-$ reconstruction at the Z-pole}
\begin{figure}[htbp]
\centering
\includegraphics[width=.45\textwidth]{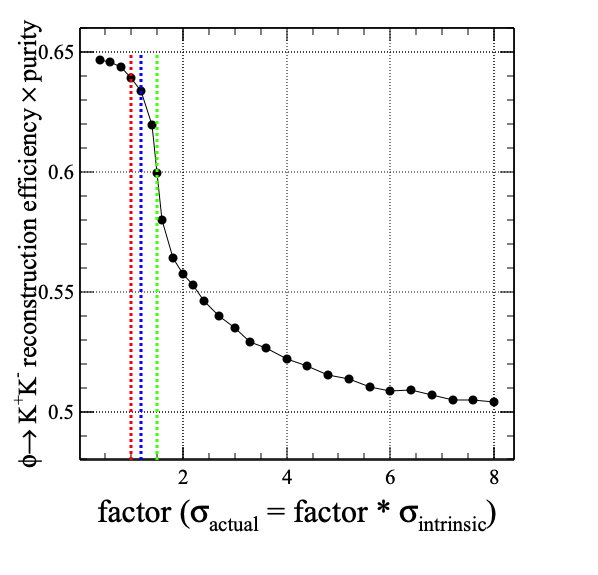}
\caption{\label{resoPhi}The distribution of $\phi\to K^+K^-$ reconstruction as a function of the factor defined in $\sigma_{actual} = factor \cdot \sigma_{intrinsic}$. The red/blue/green line corresponds to the 0\%/20\%/50\% degradation of the dE/dx resolution.}
\end{figure}

The $\phi$ plays a central role in the reconstruction of many high-level objects, especially at $B_s\to \phi\nu\overline{\nu}$, where $\phi$ is the only visible component.
This subsection investigates the performance of $\phi\to K^+K^-$ reconstruction with the inclusive hadronic Z-pole samples.
The reconstruction chain is given below.
\begin{enumerate}
\item Match all pairs of reconstructed oppositely charged tracks.
\item Set the absolute difference between the invariant mass of two tracks and the $\phi$ mass to less than $8\, $MeV/c$^2$.
\item Use the kinematic fitting package to reconstruct the vertex of the two matched tracks and set the value of the vertex $\chi^2$ to less than 9.95.
\item The identified positively charged particle must be a $K^+$ and a negative $K^-$.
\end{enumerate}
Notably, $\phi$ could come from QCD directly or from a decay of other hadrons.
Since there is no requirement to identify these two sources, the selection conditions are not applied in a further step.
The event selection process is shown in Table~\ref{phi}, where the definition of efficiency and purity is similar to that in the $D^0$ reconstruction.
The effects of dE/dx resolution on $\phi$ reconstruction performance are shown in Fig.~\ref{resoPhi}.

\begin{table}[htbp]
\centering
\caption{\label{phi}The efficiency and purity of $\phi\to K^+K^-$ reconstructed using Z-pole sample. The second column ($\epsilon$) lists selection efficiency after each selection step, and the third column (p) lists the corresponding purity.}
\smallskip
\begin{tabular}{ccc}
\hline
 										        	&$\epsilon$(\%)			& p (\%)\\
 \hline
$|mass - mass_{D0}| < 8\, $MeV/c$^2 $                      & $89.42 \pm 0.07$            &  $6.43 \pm 0.06$            \\
vertex fitted $\chi^2 <  9.95$ 		               & $83.15 \pm 0.09$             &  $9.14 \pm 0.07  $           \\
PID                                                &  $82.26 \pm 0.09$           &  $77.70 \pm 0.10$                       \\
\hline
\end{tabular}
\end{table}

\section{Conclusion}
\label{sec:Con}

The Z-pole mode of operation of the CEPC provides a great opportunity for flavor physics, where PID performance is essential.
The PID studied in this article is derived from the TOF and dE/dx information collected by ECAL and TPC of the CEPC baseline detector.
Using a GEANT4-based MC simulation, the dE/dx resolution can be better than 2.5\% in the barrel region with an momentum of the incident charged particle larger than $2\, $GeV/c.
In a real experiment, the dE/dx resolution would be degraded by the effects of the detector and electronic readout.
At CEPC, dE/dx resolution is pursued to be better than 3\% in the barrel region, which corresponds to 1/5 degradation of the intrinsic dE/dx resolution.

With inclusive hadronic Z-pole samples, the benchmarks of $K^{\pm}$ identification performance as well as $D^0\to \pi^+K^-$ and $\phi\to K^+K^-$ reconstruction performance is used to quantify the PID performance of CEPC.
The $K^{\pm}$ identification efficiency and purity can reach 95.97\% and 81.56\%, respectively, only with dE/dx information.
After including the TOF information, the $K^{\pm}$ identification efficiency and purity can be improved by 2.56\% and 20.02\%, respectively.
With the degradation of the dE/dx resolution by less than 20\%, the $K^{\pm}$ identification efficiency/purity can be better than 97\%/96\%.
The $D^0\to \pi^+K^-$($\phi\to K^+K^-$) reconstruction performance also relies strongly on the PID performance and can be reconstructed with efficiency/purity of 68.19\%/89.05\% (82.26\%/77.70\%).
The PID performance in the CEPC's Z-pole operating mode provides solid support for relevant flavor physics measurements.

The analyses described in this article are used to evaluate PID performance and provide a reference for TPC optimization.
The PID performance of the TPC could be further improved through the use of a cluster counting method (dN/dx).

\section*{Acknowledgment}
We thank Fenfen An, Taifan Zheng, Zhiyang Yuan, Yuexing Wang, and Yuzhi Che for their supports and helps.
We thank Gang Li and Chengdong Fu for producing the samples. 
This project is supported by the International Partnership Program of Chinese Academy of Sciences (Grant No. 113111KYSB20190030), the Innovative Scientific Program of Institute of High Energy Physics.


\bibliographystyle{unsrt}

\end{document}